\documentclass[12pt]{article}
\begin{document}

\author{C.~Bizdadea\thanks{%
e-mail address: bizdadea@central.ucv.ro}, M.~T.~Miaut\u {a}, S.~O.~Saliu%
\thanks{%
e-mail address: osaliu@central.ucv.ro} \\
Faculty of Physics, University of Craiova\\
13 A.~I.~Cuza Str., Craiova RO-1100, Romania}
\title{Hamiltonian BRST interactions in Abelian theories}
\maketitle

\begin{abstract}
Consistent couplings between an Abelian gauge field and three types of
matter fields are investigated by means of the Hamiltonian BRST deformation
theory based on cohomological techniques. In this manner, scalar
electrodynamics, the Stuckelberg theory for Abelian zero- and one-forms,
respectively, spinor electrodynamics, are inferred.

\end{abstract}

\section{Introduction}

The reformulation of the Lagrangian BRST symmetry~\cite{1,2,3,4,5} on
cohomological grounds allowed, among others, the study of consistent
interactions that can be introduced among fields with gauge freedom without
changing the number of gauge symmetries~\cite{6,7,8,9,10} with the help of
the deformation of the master equation~\cite{11} in the framework of the
local BRST cohomology~\cite{11,13,14,15,16,16a}. This Lagrangian
cohomological deformation technique has been successfully applied to many
models of interest, like Chern-Simons models, Yang-Mills theories, the
Chapline-Manton model, $p$-forms and chiral $p$-forms, Einstein's gravity
theory, four- and eleven-dimensional supergravity, or BF models~\cite{11},~%
\cite{17,18,19,20,21,22,23,24,25,26,27,28,29,30,30a,30b}.

On the other hand, the Hamiltonian BRST formalism~\cite{5},~\cite
{31,32,33,34,35} appears to be the most natural setting for implementing the
BRST symmetry in quantum mechanics~\cite{5} (Chapter 14). In the meantime,
it attracted much attention by providing a strong tool for examining
anomalies~\cite{36}, computing local BRST cohomologies~\cite{37}, as well as
for establishing a proper connection with canonical quantization formalisms,
like, for instance, the reduced phase-space or Dirac quantization procedures~%
\cite{38}. Lately, the Hamiltonian BRST approach has been extended to the
investigation of consistent interactions that can be added in gauge theories
with the help of the deformation technique based on local cohomologies~\cite
{39,40,41,42}.

In this paper we investigate the consistent Hamiltonian interactions that
can be introduced between an Abelian gauge field and three types of matter
fields, namely, the complex scalar, the massless real scalar and Dirac, with
the help of cohomological BRST arguments combined with the deformation
technique. In each of the three cases under consideration we start from a
``free'' theory, whose Lagrangian action is equal to the sum of the action
of an Abelian gauge field and the one describing one of the matter fields.
Every of the ``free'' systems displays two types of symmetries: a rigid one
related to the matter component, that induces a certain conserved current,
and the other purely gauge, characteristic to Maxwell's theory. The
Hamiltonian BRST symmetry of the ``free'' models, $s$, simply decomposes
into $s=\delta +\gamma $, with $\delta $ the Koszul-Tate differential and $%
\gamma $ the exterior derivative along the gauge orbits. Its non-trivial
action is essentially due to the first-class constraints of the
electromagnetic field. It has been shown in~\cite{39,40,41,42} that the
Hamiltonian problem of introducing consistent interactions in gauge theories
can be reformulated as a deformation problem of the BRST charge and
BRST-invariant Hamiltonian of a starting `free' theory. Following this line,
we prove that the deformed BRST charge consistent at all orders in the
deformation parameter can be taking non-vanishing only at order one in the
case of all the investigated models. Meanwhile, the first-order deformation
of the BRST charge reduces every time to the component of antighost number
zero, which is $\gamma $-invariant. Further, we solve the equations
responsible for the deformation of the BRST-invariant Hamiltonian associated
with the ``free'' systems. Related to the first-order deformation equation
written in a local form, we give evidence for its relationship with the
conserved currents corresponding to some rigid transformations of the matter
fields. On account of this relationship, we can determine the deformed BRST
charge and the first-order deformation of the BRST-invariant Hamiltonian.
The remaining higher-order equations are then satisfactorily solved, and the
deformed BRST-invariant Hamiltonian is completely output in every of the
cases under study. It is important to notice that there appear no
obstructions regarding the locality of the deformed BRST quantities.
Analyzing the resulting interacting models, it follows that we have
constructed nothing but the scalar electrodynamics, the Stuckelberg theory
for Abelian zero- and one-forms and the spinor electrodynamics. The matter
fields are endowed, as a consequence of their couplings to the Abelian gauge
field, with some gauge transformations that can be inferred from the
original global ones merely by gauging.

This paper is organized in five sections. Section 2 briefly formulates the
analysis of consistent Hamiltonian interactions that can be added to a
``free'' theory without changing its number of gauge symmetries as a
deformation problem of the corresponding BRST charge and BRST-invariant
Hamiltonian, finally expressed in terms of the so-called main equations.
Based on this, in Section 3 we construct the consistent Hamiltonian
couplings between an Abelian gauge field and a scalar field on cohomological
grounds. As a consequence, we infer the scalar electrodynamics in the
complex case, respectively, the Stuckelberg model involving zero- and
one-forms in the massless real case. Section 4 deals with a similar topic
with respect to an Abelian gauge field and a Dirac field, leading to spinor
electrodynamics. Section 5 ends the paper with some conclusions.

\section{Main equations of the Hamiltonian BRST deformation procedure}

We consider a ``free'' Lagrangian theory, whose action is invariant under
some gauge transformations, that can in principle be reducible. At the
Lagrangian level, all the information on the gauge structure and
reducibility relations is encoded within the solution to the master
equation. Moreover, it has been shown that the deformation of the solution
to the master equation generates consistent interactions among fields with
gauge freedom~\cite{5}. At the Hamiltonian level, the gauge structure of a
given gauge theory is completely captured by the BRST charge and
BRST-invariant Hamiltonian. Similarly to the Lagrangian deformation
procedure, we can reformulate the problem of introducing consistent
Hamiltonian interactions like a deformation problem of the BRST charge and
BRST-invariant Hamiltonian.

If the interactions can be consistently constructed, then the BRST charge of
a given ``free'' theory, $\Omega _{0}$, can be deformed as 
\begin{eqnarray}
&&\Omega _{0}\rightarrow \Omega =\Omega _{0}+g\int \mathrm{d}^{D-1}x\,\omega
_{1}+g^{2}\int \mathrm{d}^{D-1}x\,\omega _{2}+O\left( g^{3}\right) = 
\nonumber \\
&&\Omega _{0}+g\Omega _{1}+g^{2}\Omega _{2}+O\left( g^{3}\right) ,
\label{s1}
\end{eqnarray}
where $\Omega $ should satisfy the equation 
\begin{equation}
\left[ \Omega ,\Omega \right] =0.  \label{s2}
\end{equation}
Here, the symbol $\left[ ,\right] $ denotes either the Poisson, or the Dirac
bracket. If the initial system is purely first-class, we need the Poisson
bracket; if there are also second-class constraints, then we eliminate them,
and work with the Dirac one. Equation (\ref{s2}) splits accordingly the
deformation parameter $g$ as 
\begin{equation}
\left[ \Omega _{0},\Omega _{0}\right] =0,  \label{s3}
\end{equation}
\begin{equation}
2\left[ \Omega _{0},\Omega _{1}\right] =0,  \label{s4}
\end{equation}
\begin{equation}
2\left[ \Omega _{0},\Omega _{2}\right] +\left[ \Omega _{1},\Omega
_{1}\right] =0,  \label{s5}
\end{equation}
\[
\vdots 
\]
Obviously, (\ref{s3}) is automatically satisfied. From the remaining
equations we deduce the pieces $\left( \Omega _{k}\right) _{k>0}$ on account
of the ``free'' BRST differential. With the deformed BRST charge at hand, we
then deform the BRST-invariant Hamiltonian of the ``free'' theory, $H_{0B}$,
like 
\begin{eqnarray}
&&H_{0B}\rightarrow H_{B}=H_{0B}+g\int \mathrm{d}^{D-1}x\,h_{1}+g^{2}\int 
\mathrm{d}^{D-1}x\,h_{2}+O\left( g^{3}\right) =  \nonumber \\
&&H_{0B}+gH_{1}+g^{2}H_{2}+O\left( g^{3}\right) ,  \label{s6}
\end{eqnarray}
and require that 
\begin{equation}
\left[ H_{B},\Omega \right] =0.  \label{s7}
\end{equation}
Equation (\ref{s7}) can be analyzed order by order in the deformation
parameter $g$, leading to 
\begin{equation}
\left[ H_{0B},\Omega _{0}\right] =0,  \label{s8}
\end{equation}
\begin{equation}
\left[ H_{0B},\Omega _{1}\right] +\left[ H_{1},\Omega _{0}\right] =0,
\label{s9}
\end{equation}
\begin{equation}
\left[ H_{0B},\Omega _{2}\right] +\left[ H_{1},\Omega _{1}\right] +\left[
H_{2},\Omega _{0}\right] =0,  \label{s10}
\end{equation}
\[
\vdots 
\]
Clearly, (\ref{s8}) is again fulfilled, while from the others one can
determine the components $\left( H_{k}\right) _{k>0}$ by relying on the BRST
symmetry of the ``free'' model. Equations (\ref{s3}-- \ref{s5}), \textit{etc.%
}, and (\ref{s8}--\ref{s10}), \textit{etc.}, represent the main equations of
our Hamiltonian deformation procedure. They will be solved in the next
sections with respect to the models under study by means of some
cohomological techniques, specific to the Hamiltonian BRST formalism.

\section{Couplings between an Abelian gauge field and a scalar field}

Initially, we investigate the consistent Hamiltonian couplings between an
Abelian gauge field and a scalar field along the line exposed in the above,
and derive the scalar electrodynamics in the complex case, respectively, the
Stuckelberg coupling in the massless real case.

\subsection{Couplings with a complex scalar field}

We start from a ``free'' Lagrangian action written as the sum between the
action of an Abelian gauge field $A_{\mu }$ and that of a complex scalar
field $\left( \varphi ,\bar{\varphi}\right) $%
\begin{equation}
S_{0}^{L}\left[ \varphi ,\bar{\varphi},A^{\mu }\right] =\int \mathrm{d}%
^{4}x\left( -\frac{1}{4}F_{\mu \nu }F^{\mu \nu }+\left( \partial _{\mu
}\varphi \right) \partial ^{\mu }\bar{\varphi}-\mu ^{2}\varphi \bar{\varphi}%
-V\left( \varphi \bar{\varphi}\right) \right) ,  \label{s11}
\end{equation}
where the bar operation represents the complex conjugation, while the
Abelian field strength is defined in the usual manner by $F_{\mu \nu
}=\partial _{\mu }A_{\nu }-\partial _{\nu }A_{\mu }$.

As commonly known, the action (\ref{s11}) is invariant under the
one-parameter rigid symmetries (genuinely, only the part corresponding to
the scalar field is non-trivially responsible for this global invariance) 
\begin{equation}
\Delta \varphi =i\varphi \xi ,\;\Delta \bar{\varphi}=-i\bar{\varphi}\xi ,
\label{s12}
\end{equation}
leading, via Noether's theorem, to the conservation law 
\begin{equation}
\partial _{\mu }j^{\mu }=i\varphi \frac{\delta \mathcal{L}^{\left( S\right) }%
}{\delta \varphi }-i\bar{\varphi}\frac{\delta \mathcal{L}^{\left( S\right) }%
}{\delta \bar{\varphi}},  \label{s13}
\end{equation}
giving evidence for the conserved current 
\begin{equation}
j^{\mu }=i\left( \bar{\varphi}\partial ^{\mu }\varphi -\varphi \partial
^{\mu }\bar{\varphi}\right) ,  \label{s14}
\end{equation}
with 
\begin{equation}
\frac{\delta \mathcal{L}^{\left( S\right) }}{\delta \varphi }=-\left(
\partial _{\mu }\partial ^{\mu }+\mu ^{2}+\frac{\partial V}{\partial \left(
\varphi \bar{\varphi}\right) }\right) \bar{\varphi},  \label{s15}
\end{equation}
\begin{equation}
\frac{\delta \mathcal{L}^{\left( S\right) }}{\delta \bar{\varphi}}=-\left(
\partial _{\mu }\partial ^{\mu }+\mu ^{2}+\frac{\partial V}{\partial \left(
\varphi \bar{\varphi}\right) }\right) \varphi ,  \label{s16}
\end{equation}
where $\mathcal{L}^{\left( S\right) }$ stands for the Lagrangian density
associated with the complex scalar field.

By passing to the canonical analysis of the action (\ref{s11}), we find the
Abelian first-class constraints and first-class Hamiltonian of the form 
\begin{equation}
G_{1}\equiv \pi _{0}\approx 0,\;G_{2}\equiv -\partial ^{i}\pi _{i}\approx 0,
\label{s17}
\end{equation}
\begin{eqnarray}
&&H_{0}=\int \mathrm{d}^{3}x\left( \frac{1}{2}\pi _{i}\pi _{i}+\frac{1}{4}%
F_{ij}F^{ij}-A^{0}\partial ^{i}\pi _{i}+\right.  \nonumber \\
&&\left. \pi \bar{\pi}-\left( \partial _{j}\varphi \right) \partial ^{j}\bar{%
\varphi}+\mu ^{2}\varphi \bar{\varphi}+V\left( \varphi \bar{\varphi}\right)
\right) ,  \label{s19}
\end{eqnarray}
where $\pi _{\mu }$, $\pi $ and $\bar{\pi}$ denote the canonical momenta of
the fields $A^{\mu }$, $\varphi $, respectively, $\bar{\varphi}$. The BRST
charge of this ``free'' theory is then 
\begin{equation}
\Omega _{0}=\int \mathrm{d}^{3}x\left( \pi _{0}\eta ^{1}-\left( \partial
^{i}\pi _{i}\right) \eta ^{2}\right) ,  \label{s21}
\end{equation}
where $\eta ^{1}$and $\eta ^{2}$\ represent the fermionic Hamiltonian
ghosts. Their antighosts, to be denoted by $\mathcal{P}_{1}$, respectively, $%
\mathcal{P}_{2}$, are also fermionic. The ``free'' Hamiltonian BRST symmetry 
$s\bullet =\left[ \bullet ,\Omega _{0}\right] $ simply decomposes as 
\begin{equation}
s=\delta +\gamma ,  \label{s22}
\end{equation}
with $\delta $ the Koszul-Tate differential, and $\gamma $ the exterior
longitudinal derivative along the gauge orbits. The Koszul-Tate differential
is graded accordingly the antighost number ($\mathrm{antigh}$, $\mathrm{%
antigh}(\delta )=-1$), the degree of the exterior longitudinal derivative is
named pure ghost number ($\mathrm{pgh}$, $\mathrm{pgh}(\gamma )=1$, $\mathrm{%
pgh}\left( \delta \right) =0$, $\mathrm{antigh}(\gamma )=0$), while the
overall grading of the BRST differential is called ghost number ($\mathrm{gh}
$, $\mathrm{gh}\left( s\right) =1$), and is defined by the difference
between the pure ghost number and the antighost number. The degrees of the
generators from the BRST complex are valued 
\begin{equation}
\mathrm{antigh}\left( A^{\mu }\right) =\mathrm{antigh}\left( \pi _{\mu
}\right) =\mathrm{antigh}\left( \varphi \right) =\mathrm{antigh}\left( \bar{%
\varphi}\right) =0,  \label{s22a}
\end{equation}
\begin{equation}
\mathrm{antigh}\left( \pi \right) =\mathrm{antigh}\left( \bar{\pi}\right)
=0,\;\mathrm{antigh}\left( \mathcal{P}_{a}\right) =1,\;\mathrm{antigh}\left(
\eta ^{a}\right) =0,\;a=1,2,  \label{s22b}
\end{equation}
\begin{equation}
\mathrm{pgh}\left( A^{\mu }\right) =\mathrm{pgh}\left( \pi _{\mu }\right) =%
\mathrm{pgh}\left( \varphi \right) =\mathrm{pgh}\left( \bar{\varphi}\right) =%
\mathrm{pgh}\left( \pi \right) =\mathrm{pgh}\left( \bar{\pi}\right) =0,
\label{s22c}
\end{equation}
\begin{equation}
\mathrm{pgh}\left( \mathcal{P}_{a}\right) =0,\;\mathrm{pgh}\left( \eta
^{a}\right) =1,\;a=1,2.  \label{s22d}
\end{equation}
The operators $\delta $ and $\gamma $ act on the BRST generators through the
relations 
\begin{equation}
\delta A^{\mu }=0,\;\delta \pi _{\mu }=0,\;\delta \varphi =0,\;\delta \bar{%
\varphi}=0,\;\delta \pi =0,  \label{s23}
\end{equation}
\begin{equation}
\delta \bar{\pi}=0,\;\delta \mathcal{P}_{1}=-\pi _{0},\;\delta \mathcal{P}%
_{2}=\partial ^{i}\pi _{i},\;\delta \eta ^{1}=0,\;\delta \eta ^{2}=0,
\label{s24}
\end{equation}
\begin{equation}
\gamma A^{0}=\eta ^{1},\;\gamma A^{i}=\partial ^{i}\eta ^{2},\;\gamma \pi
_{\mu }=0,\;\gamma \varphi =0,\;\gamma \bar{\varphi}=0,\;\gamma \pi =0,
\label{s25}
\end{equation}
\begin{equation}
\gamma \bar{\pi}=0,\;\gamma \mathcal{P}_{1}=0,\;\gamma \mathcal{P}%
_{2}=0,\;\gamma \eta ^{1}=0,\;\gamma \eta ^{2}=0,  \label{s26}
\end{equation}
that will be used in the sequel at the deformation procedure.

Next, we solve the equations (\ref{s4}--\ref{s5}), \textit{etc.}, and (\ref
{s9}--\ref{s10}), \textit{etc.}, that govern the Hamiltonian deformation.
Taking into account the expression (\ref{s2}), the local form of (\ref{s4})
holds if and only if $\omega _{1}$ is an $s$-co-cycle modulo the spatial
part of the space-time derivative, $\tilde{\mathrm{d}}=\mathrm{%
dx^{i}\partial _{i}}$, hence if and only if 
\begin{equation}
s\omega _{1}=\partial _{k}\sigma ^{k},  \label{s27}
\end{equation}
for some $\sigma ^{k}$. In order to solve (\ref{s27}) we expand $\omega _{1}$
according to the antighost number 
\begin{equation}
\omega _{1}=\stackrel{\left( 0\right) }{\omega }_{1}+\stackrel{\left(
1\right) }{\omega }_{1}+\cdots +\stackrel{\left( J\right) }{\omega }_{1},
\label{s28}
\end{equation}
where the last term can be assumed to be annihilated by $\gamma $. As $%
\mathrm{antigh}\left( \stackrel{\left( J\right) }{\omega }_{1}\right) =J$
and $\mathrm{gh}\left( \stackrel{\left( J\right) }{\omega }_{1}\right) =1$,
we find the result that $\mathrm{pgh}\left( \stackrel{\left( J\right) }{%
\omega }_{1}\right) =J+1$, so we can represent $\stackrel{\left( J\right) }{%
\omega }_{1}$ in the form $\stackrel{\left( J\right) }{\omega }_{1}=\mu
_{J}\left( \eta ^{2}\right) ^{J+1}$. (The ghost $\eta ^{1}$ does not come
into discussion as it is trivial in the cohomology of $\gamma $: $\gamma
A^{0}=\eta ^{1}$, $\gamma \eta ^{1}=0$.) Due to the fermionic character of $%
\eta ^{2}$, this term is non-vanishing if and only if $J=0$, such that 
\begin{equation}
\omega _{1}=\stackrel{\left( 0\right) }{\omega }_{1}=\mu _{0}\eta ^{2}.
\label{s29}
\end{equation}
With this choice, it is easy to check that the $\gamma $-invariant
coefficient $\mu _{0}$ should satisfy the conditions $\mathrm{antigh}\left(
\mu _{0}\right) =0$, $\mathrm{pgh}\left( \mu _{0}\right) =0$, $\gamma \mu
_{0}=0$. From (\ref{s23}--\ref{s26}) the result is obtained that $\mu _{0}$
can depend on $\left( \varphi ,\bar{\varphi}\right) $ and $\left( \pi ,\bar{%
\pi}\right) $, so $\mu _{0}=\mu _{0}\left( \varphi ,\bar{\varphi},\pi ,\bar{%
\pi}\right) $. In this way, the first-order deformation of the BRST charge,
determined up to $\mu _{0}$, is given by 
\begin{equation}
\Omega _{1}=\int \mathrm{d}^{3}x\,\mu _{0}\left( \varphi ,\bar{\varphi},\pi ,%
\bar{\pi}\right) \eta ^{2}.  \label{s30}
\end{equation}
By direct computation we then obtain that $\left[ \Omega _{1},\Omega
_{1}\right] =0$, no matter what $\mu _{0}\left( \varphi ,\bar{\varphi},\pi ,%
\bar{\pi}\right) $ we take. The second-order deformation equation of the
BRST charge, (\ref{s5}), is thus satisfied for $\Omega _{2}=0$,\ such that
the corresponding higher-order deformations can be taken as $\Omega
_{3}=\Omega _{4}=\cdots =0$. Consequently, the overall deformed BRST charge
takes the form 
\begin{equation}
\Omega =\int \mathrm{d}^{3}x\left( \pi _{0}\eta ^{1}-\left( \partial ^{i}\pi
_{i}-g\mu _{0}\left( \varphi ,\bar{\varphi},\pi ,\bar{\pi}\right) \right)
\eta ^{2}\right) .  \label{s31}
\end{equation}

At this point, we investigate the deformation of the BRST-invariant
Hamiltonian, described by (\ref{s9}--\ref{s10}), \textit{etc.}, where the
BRST-invariant Hamiltonian of the free theory reads 
\begin{equation}
H_{0B}=H_{0}+\int \mathrm{d}^{3}x\,\eta ^{1}\mathcal{P}_{2}.  \label{s32}
\end{equation}
From (\ref{s19}), (\ref{s30}) and (\ref{s32}) we see that the first term in (%
\ref{s9}) can be written as 
\begin{eqnarray}
&&\left[ H_{0B},\Omega _{1}\right] =\int \mathrm{d}^{3}x\left( -\left( \pi 
\bar{u}+\bar{\pi}u\right) \eta ^{2}-\eta ^{1}\mu _{0}+\right.  \nonumber \\
&&\left. \left( \left( \partial _{j}\partial ^{j}\varphi +\mu ^{2}\varphi +%
\frac{\partial V}{\partial \left( \varphi \bar{\varphi}\right) }\varphi
\right) \bar{v}+\left( \partial _{j}\partial ^{j}\bar{\varphi}+\mu ^{2}\bar{%
\varphi}+\frac{\partial V}{\partial \left( \varphi \bar{\varphi}\right) }%
\bar{\varphi}\right) v\right) \eta ^{2}\right)  \nonumber \\
&=&\int \mathrm{d}^{3}x\,\lambda ,  \label{s33}
\end{eqnarray}
so that the local form of (\ref{s9}) leads to 
\begin{equation}
sh_{1}+\lambda =\partial ^{i}n_{i},  \label{s34}
\end{equation}
for some $n_{i}$. In the above we used the notations $\bar{u}\left( x\right)
=\int \mathrm{d}^{3}y\frac{\delta \mu _{0}\left( x^{0},\vec{y}\right) }{%
\delta \bar{\varphi}\left( x\right) }$, $u\left( x\right) =\int \mathrm{d}%
^{3}y\frac{\delta \mu _{0}\left( x^{0},\vec{y}\right) }{\delta \varphi
\left( x\right) }$, $\bar{v}\left( x\right) =\int \mathrm{d}^{3}y\frac{%
\delta \mu _{0}\left( x^{0},\vec{y}\right) }{\delta \bar{\pi}\left( x\right) 
}$, $v\left( x\right) =\int \mathrm{d}^{3}y\frac{\delta \mu _{0}\left( x^{0},%
\vec{y}\right) }{\delta \pi \left( x\right) }$. As the term $-\eta ^{1}\mu
_{0}$ from $\lambda $ does not contain spatial derivatives, it should be
compensated by a similar term of opposite sign in $sh_{1}$. This can be
achieved if and only if 
\begin{equation}
h_{1}=\mu _{0}A^{0}+\alpha ,  \label{s35}
\end{equation}
where $\alpha $ should depend on $A^{i}$ in order to produce a term
containing spatial derivatives through its Poisson bracket with the second
term in (\ref{s21}). In the meantime, $\alpha $ involves no ghosts or
antighosts because otherwise we would enlarge $\left[ H_{1},\Omega
_{0}\right] $ with pieces that are not present in $\left[ H_{0B},\Omega
_{1}\right] $. These considerations further give 
\begin{equation}
\left[ H_{1},\Omega _{0}\right] =\int \mathrm{d}^{3}x\left( \eta ^{1}\mu
_{0}+a_{i}\partial ^{i}\eta ^{2}\right) ,  \label{s36}
\end{equation}
which combined with (\ref{s33}), lead to the concrete form of (\ref{s34}) as 
\begin{eqnarray}
&&\left( \left( \partial _{j}\partial ^{j}\varphi +\mu ^{2}\varphi +\frac{%
\partial V}{\partial \left( \varphi \bar{\varphi}\right) }\varphi \right) 
\bar{v}+\left( \partial _{j}\partial ^{j}\bar{\varphi}+\mu ^{2}\bar{\varphi}+%
\frac{\partial V}{\partial \left( \varphi \bar{\varphi}\right) }\bar{\varphi}%
\right) v-\right.  \nonumber \\
&&\left. \pi \bar{u}-\bar{\pi}u\right) \eta ^{2}+a_{i}\partial ^{i}\eta
^{2}=\partial ^{i}n_{i},  \label{s37}
\end{eqnarray}
where $a_{i}\left( x\right) =\int \mathrm{d}^{3}y\frac{\delta \alpha \left(
x^{0},\vec{y}\right) }{\delta A^{i}\left( x\right) }$. In order to obtain a
total derivative in the left-hand side of (\ref{s37}) we must have 
\begin{eqnarray}
&&\left( \partial _{j}\partial ^{j}\varphi +\mu ^{2}\varphi +\frac{\partial V%
}{\partial \left( \varphi \bar{\varphi}\right) }\varphi \right) \bar{v}%
+\left( \partial _{j}\partial ^{j}\bar{\varphi}+\mu ^{2}\bar{\varphi}+\frac{%
\partial V}{\partial \left( \varphi \bar{\varphi}\right) }\bar{\varphi}%
\right) v-  \nonumber \\
&&\pi \bar{u}-\bar{\pi}u=\partial ^{i}a_{i}.  \label{s38}
\end{eqnarray}
By adding the terms $\left( \partial _{0}\partial ^{0}\varphi \right) \bar{v}
$ and $\left( \partial _{0}\partial ^{0}\bar{\varphi}\right) v$ to both
hand-sides of the above equation, we arrive at 
\begin{eqnarray}
&&-\frac{\delta \mathcal{L}^{\left( S\right) }}{\delta \varphi }v-\frac{%
\delta \mathcal{L}^{\left( S\right) }}{\delta \bar{\varphi}}\bar{v}=\pi \bar{%
u}+\bar{\pi}u+  \nonumber \\
&&\left( \partial _{0}\partial ^{0}\varphi \right) \bar{v}+\left( \partial
_{0}\partial ^{0}\bar{\varphi}\right) v+\partial ^{i}a_{i}.  \label{s39}
\end{eqnarray}
The left-hand side of (\ref{s39}) represents nothing but the variation of
the Lagrangian density of the complex scalar field under the rigid
transformations 
\begin{equation}
\Delta \varphi \left( x\right) =-\int \mathrm{d}^{3}y\frac{\delta \mu
_{0}\left( x^{0},\vec{y}\right) }{\delta \pi \left( x\right) }\xi ,\;\Delta 
\bar{\varphi}\left( x\right) =-\int \mathrm{d}^{3}y\frac{\delta \mu
_{0}\left( x^{0},\vec{y}\right) }{\delta \bar{\pi}\left( x\right) }\xi .
\label{s40}
\end{equation}
On the other hand, by identifying the above global variations with the rigid
one-parameter transformations (\ref{s12}), we get the equations 
\begin{equation}
\int \mathrm{d}^{3}y\frac{\delta \mu _{0}\left( x^{0},\vec{y}\right) }{%
\delta \pi \left( x\right) }=-i\varphi \left( x\right) ,\;\int \mathrm{d}%
^{3}y\frac{\delta \mu _{0}\left( x^{0},\vec{y}\right) }{\delta \bar{\pi}%
\left( x\right) }=i\bar{\varphi}\left( x\right) ,  \label{s41}
\end{equation}
whose solution outputs the unknown function $\mu _{0}$ of the type 
\begin{equation}
\mu _{0}\left( y\right) =i\left( \bar{\varphi}\bar{\pi}-\varphi \pi \right)
\left( y\right) .  \label{s42}
\end{equation}
Inserting (\ref{s42}) in (\ref{s39}), and taking into account (\ref{s13}),
we find that $\int \mathrm{d}^{3}y\frac{\delta \alpha \left( x^{0},\vec{y}%
\right) }{\delta A^{i}\left( x\right) }=i\left( \bar{\varphi}\partial
_{i}\varphi -\varphi \partial _{i}\bar{\varphi}\right) \left( x\right) $,
which yields 
\begin{equation}
\alpha \left( y\right) =\left( i\left( \bar{\varphi}\partial _{i}\varphi
-\varphi \partial _{i}\bar{\varphi}\right) A^{i}\right) \left( y\right) .
\label{s43}
\end{equation}
In this manner, we have completely determined the first-order deformation of
the BRST-invariant Hamiltonian and BRST charge: 
\begin{equation}
H_{1}=i\int \mathrm{d}^{3}x\left( \left( \bar{\varphi}\bar{\pi}-\varphi \pi
\right) A^{0}+\left( \bar{\varphi}\partial _{i}\varphi -\varphi \partial _{i}%
\bar{\varphi}\right) A^{i}\right) ,  \label{s44}
\end{equation}
\begin{equation}
\Omega _{1}=i\int \mathrm{d}^{3}x\left( \bar{\varphi}\bar{\pi}-\varphi \pi
\right) \eta ^{2}.  \label{sx44}
\end{equation}

Next, we approach the equation responsible for the second-order deformation
of the BRST-invariant Hamiltonian, (\ref{s10}). In view of this, we remark
that the first term is vanishing as $\Omega _{2}=0$, while the second one is
equal to 
\begin{equation}
\left[ H_{1},\Omega _{1}\right] =-2\int \mathrm{d}^{3}x\left( \partial
_{i}\left( \varphi \bar{\varphi}A^{i}\right) \right) \eta ^{2}=\int \mathrm{d%
}^{3}x\,\rho .  \label{sy44}
\end{equation}
Consequently, (\ref{s10}) written in a local form becomes 
\begin{equation}
sh_{2}+\rho =\partial _{i}k^{i},  \label{sz44}
\end{equation}
whose solution reads 
\begin{equation}
h_{2}=-\varphi \bar{\varphi}A^{i}A_{i},  \label{sw44}
\end{equation}
such that 
\begin{equation}
sh_{2}+\rho =\partial _{i}\left( -2\varphi \bar{\varphi}A^{i}\eta
^{2}\right) .  \label{sk44}
\end{equation}
Passing now to the third-order equation, $\left[ H_{0B},\Omega _{3}\right]
+\left[ H_{1},\Omega _{2}\right] +\left[ H_{2},\Omega _{1}\right] +\left[
H_{3},\Omega _{0}\right] =0$, we remark that the first two terms vanish as $%
\Omega _{2}=\Omega _{3}=0$, while by direct computation we obtain 
\begin{equation}
\left[ H_{2},\Omega _{1}\right] =0.  \label{sl44}
\end{equation}
Thus, we can safely take the third-order deformation piece in the
BRST-invariant Hamiltonian to be equal to zero, $H_{3}=0$, and, moreover, it
turns out that all higher-order deformation equations are fulfilled for 
\begin{equation}
H_{4}=H_{5}=\cdots =0.  \label{s45}
\end{equation}

Synthesizing the results deduced so far, we find that the complete
deformations of the BRST charge and BRST-invariant Hamiltonian associated
with the ``free'' system under discussion are 
\begin{equation}
\Omega =\int \mathrm{d}^{3}x\left( \pi _{0}\eta ^{1}-\left( \partial ^{i}\pi
_{i}-ig\left( \bar{\varphi}\bar{\pi}-\varphi \pi \right) \right) \eta
^{2}\right) ,  \label{s46}
\end{equation}
respectively, 
\begin{eqnarray}
&&H_{B}=\int \mathrm{d}^{3}x\left( \frac{1}{2}\pi _{i}\pi _{i}+\frac{1}{4}%
F_{ij}F^{ij}-A^{0}\partial ^{i}\pi _{i}+\pi \bar{\pi}-\right.  \nonumber \\
&&\left( \partial _{j}\varphi \right) \left( \partial ^{j}\bar{\varphi}%
\right) +\mu ^{2}\varphi \bar{\varphi}+V\left( \varphi \bar{\varphi}\right)
+ig\left( \bar{\varphi}\bar{\pi}-\varphi \pi \right) A^{0}+  \nonumber \\
&&\left. ig\left( \bar{\varphi}\partial _{i}\varphi -\varphi \partial _{i}%
\bar{\varphi}\right) A^{i}-g^{2}\varphi \bar{\varphi}A^{i}A_{i}+\eta ^{1}%
\mathcal{P}_{2}\right) .  \label{s47}
\end{eqnarray}
Now, we are in the position to analyse the resulting deformed theory. From
the pieces present in $\Omega $ that are linear in the ghosts we observe
that the resulting model displays at the Hamiltonian level the same primary
first-class constraint like the initial system (the former constraint in (%
\ref{s17})), but the secondary one as a result of the deformation process
has turned into 
\begin{equation}
\gamma _{2}\equiv -\partial ^{i}\pi _{i}+ig\left( \bar{\varphi}\bar{\pi}%
-\varphi \pi \right) \approx 0,  \label{s48}
\end{equation}
such that these first-class constraints are still Abelian. Examining the
terms that contain neither ghosts nor antighosts in (\ref{s47}), we notice
that the first-class Hamiltonian of the interacting theory reads 
\begin{eqnarray}
&&H=\int \mathrm{d}^{3}x\left( \frac{1}{2}\pi _{i}\pi _{i}+\frac{1}{4}%
F_{ij}F^{ij}-A^{0}\left( \partial ^{i}\pi _{i}-ig\left( \bar{\varphi}\bar{\pi%
}-\varphi \pi \right) \right) +\right.  \nonumber \\
&&\left. \pi \bar{\pi}-\left( D_{j}\varphi \right) \left( \overline{%
D^{j}\varphi }\right) +\mu ^{2}\varphi \bar{\varphi}+V\left( \varphi \bar{%
\varphi}\right) \right) ,  \label{s49}
\end{eqnarray}
where the spatial part of the covariant derivative is defined through 
\begin{equation}
D_{i}=\partial _{i}+igA_{i}.  \label{s50}
\end{equation}
The Lagrangian setting of the deformed system can be derived by successively
passing to the extended and total formalisms, which finally yields the
Lagrangian action 
\begin{equation}
S^{L}\left[ \varphi ,\bar{\varphi},A^{\mu }\right] =\int \mathrm{d}%
^{4}x\left( -\frac{1}{4}F_{\mu \nu }F^{\mu \nu }+\left( D_{\mu }\varphi
\right) \left( \overline{D^{\mu }\varphi }\right) -\mu ^{2}\varphi \bar{%
\varphi}-V\left( \varphi \bar{\varphi}\right) \right) ,  \label{s51}
\end{equation}
subject to the gauge transformations 
\begin{equation}
\delta _{\epsilon }A^{\mu }=\partial ^{\mu }\epsilon ,\;\delta _{\epsilon
}\varphi =ig\varphi \epsilon ,\;\delta _{\epsilon }\bar{\varphi}=-ig\bar{%
\varphi}\epsilon ,  \label{s52}
\end{equation}
with the covariant derivative given by 
\begin{equation}
D_{\mu }=\partial _{\mu }+igA_{\mu }.  \label{s53}
\end{equation}
We remark that the complex scalar field, that initially possessed only the
rigid invariances (\ref{s12}), becomes endowed now with the gauge
invariances in (\ref{s52}), that can be directly obtained from the rigid
ones merely by gauging, and, moreover, have a typical form of gauge
invariances for matter fields. It appears to be clear that the resulting
interacting theory describes, at both the Hamiltonian and the Lagrangian
level, nothing but the coupling between an Abelian gauge field and a complex
scalar field, which is known as scalar electrodynamics.

\subsection{Couplings with a massless real scalar field}

In the sequel we apply the Hamiltonian deformation scheme to a free theory
involving a massless real scalar field $\varphi $ and an Abelian gauge field 
$A^{\mu }$ , and arrive precisely at a model underlying the Stuckelberg
coupling between them. The Lagrangian action of this free system is 
\begin{equation}
S_{0}^{\prime L}\left[ \varphi ,A^{\mu }\right] =\int \mathrm{d}^{4}x\left( -%
\frac{1}{4}F_{\mu \nu }F^{\mu \nu }+\frac{1}{2}\left( \partial _{\mu
}\varphi \right) \left( \partial ^{\mu }\varphi \right) \right) ,
\label{stu1}
\end{equation}
and possesses the global shift symmetry 
\begin{equation}
\Delta \varphi =\xi ,  \label{stu2}
\end{equation}
due essentially to the presence of the real scalar field, which leads to the
conservation law 
\begin{equation}
\partial _{\mu }j^{\mu }=\frac{\delta \mathcal{L}^{(SR)}}{\delta \varphi },
\label{stu3}
\end{equation}
which reveals the conserved current 
\begin{equation}
j^{\mu }=-\partial ^{\mu }\varphi ,  \label{stu4}
\end{equation}
where 
\begin{equation}
\frac{\delta \mathcal{L}^{(SR)}}{\delta \varphi }=-\partial _{\mu }\partial
^{\mu }\varphi ,  \label{stu5}
\end{equation}
and $\mathcal{L}^{(SR)}$ denotes the Lagrangian density of the real scalar
field. From the canonical analysis of this theory we get the Abelian
first-class constraints (\ref{s17}) and the first-class Hamiltonian 
\begin{equation}
H_{0}^{\prime }=\int \mathrm{d}^{3}x\left( \frac{1}{2}\pi _{i}\pi _{i}+\frac{%
1}{4}F_{ij}F^{ij}-A^{0}\partial ^{i}\pi _{i}+\frac{1}{2}\pi ^{2}-\frac{1}{2}%
\left( \partial _{i}\varphi \right) \left( \partial ^{i}\varphi \right)
\right) ,  \label{stu6}
\end{equation}
where $\pi $ is the momentum conjugated to $\varphi $. The BRST analysis is
exactly the same like that performed for the previous model, and relies on
the formulas (\ref{s21}--\ref{s26}), from which any reference to the pair $%
\left( \bar{\varphi},\bar{\pi}\right) $ should be discarded. At this point,
we have all the elements required for the development of the Hamiltonian
deformation scheme.

The consistent deformations of the free BRST charge (\ref{s21}) demand, as
we have seen, finding the nontrivial solutions to (\ref{s4}--\ref{s5}), 
\textit{etc.} The first-order deformation equation takes the local form (\ref
{s27}). Reasoning like above, we develop $\omega _{1}$ according to the
antighost number (see (\ref{s28})), and conclude that it reduces to the
first component 
\begin{equation}
\omega _{1}=\stackrel{(0)}{\omega }_{1}=\mu _{0}^{\prime }\left( \varphi
,\pi \right) \eta ^{2},  \label{stu7}
\end{equation}
where the function $\mu _{0}^{\prime }\left( \varphi ,\pi \right) $ is
unknown and $\gamma $-invariant, such that the deformed BRST charge takes
the form (\ref{s33}), with $\mu _{0}^{\prime }\left( \varphi ,\pi \right) $
instead of $\mu _{0}\left( \varphi ,\bar{\varphi},\pi ,\bar{\pi}\right) $.

Investigating in the sequel the deformation of the BRST-invariant
Hamiltonian (\ref{s32}) (with $H_{0}$ replaced by $H_{0}^{\prime }$) at the
first-order level, described by equation (\ref{s9}), it follows, with the
help of the relation (\ref{stu7}), that 
\begin{equation}
\left[ H_{0B},\Omega _{1}\right] =\int \mathrm{d}^{3}x\left( \left( -\pi
u^{\prime }+\left( \partial _{j}\partial ^{j}\varphi \right) v^{\prime
}\right) \eta ^{2}-\eta ^{1}\mu _{0}^{\prime }\right) =\int \mathrm{d}%
^{3}x\,\lambda ^{\prime },  \label{stu8}
\end{equation}
where $u^{\prime }\left( x\right) =\int \mathrm{d}^{3}y\frac{\delta \mu
_{0}^{\prime }\left( x^{0},\vec{y}\right) }{\delta \varphi \left( x\right) }$
and $v^{\prime }\left( x\right) =\int \mathrm{d}^{3}y\frac{\delta \mu
_{0}^{\prime }\left( x^{0},\vec{y}\right) }{\delta \pi \left( x\right) }$,
hence the local form of (\ref{s9}) can be written as 
\begin{equation}
sh_{1}+\lambda ^{\prime }=\partial ^{i}n_{i}^{\prime }.  \label{stu9}
\end{equation}
Now, we take $h_{1}$ as 
\begin{equation}
h_{1}=\mu _{0}^{\prime }A^{0}+\alpha ^{\prime },  \label{stu10}
\end{equation}
in order to discard the term $\eta ^{1}\mu _{0}^{\prime }$ from the left
hand-side of (\ref{s9}), where $\alpha ^{\prime }$ has both the antighost
and pure ghost numbers equal to zero and depends in a nontrivial way of $%
A^{i}$ for the same reason as before. After some computation, we deduce that 
\begin{equation}
\left[ H_{1},\Omega _{0}\right] =\int \mathrm{d}^{3}x\left( \eta ^{1}\mu
_{0}^{\prime }+a_{i}^{\prime }\partial ^{i}\eta ^{2}\right) ,  \label{stu11}
\end{equation}
with $a_{i}^{\prime }\left( x\right) =\int \mathrm{d}^{3}y\frac{\delta
\alpha ^{\prime }\left( x^{0},\vec{y}\right) }{\delta A^{i}\left( x\right) }$%
. Therefore, (\ref{stu9}) becomes 
\begin{equation}
\left( -\pi u^{\prime }+\left( \partial _{j}\partial ^{j}\varphi \right)
v^{\prime }\right) \eta ^{2}+a_{i}^{\prime }\partial ^{i}\eta ^{2}=\partial
^{i}n_{i}^{\prime },  \label{stu12}
\end{equation}
and is satisfied if we impose 
\begin{equation}
-\pi u^{\prime }+\left( \partial _{j}\partial ^{j}\varphi \right) v^{\prime
}=\partial ^{i}a_{i}^{\prime }.  \label{stu13}
\end{equation}
If we add the term $\left( \partial _{0}\partial ^{0}\varphi \right)
v^{\prime }$ to both sides of the last equation, we find the relation 
\begin{equation}
\frac{\delta \mathcal{L}^{\left( SR\right) }}{\delta \varphi }v^{\prime
}=-\left( \pi u^{\prime }+\left( \partial _{0}\partial ^{0}\varphi \right)
v^{\prime }+\partial ^{i}a_{i}^{\prime }\right) ,  \label{stu14}
\end{equation}
whose left-hand side signifies the variation of the Lagrangian density of
the real scalar field under the one-parameter rigid transformations 
\begin{equation}
\Delta \varphi \left( x\right) =\int \mathrm{d}^{3}y\frac{\delta \mu
_{0}^{\prime }\left( x^{0},\vec{y}\right) }{\delta \pi \left( x\right) }\xi .
\label{stu15}
\end{equation}
Then, by identifying (\ref{stu15}) with the global shift invariance (\ref
{stu2}), characteristic for the real scalar field, we are led to the
equation 
\begin{equation}
\int \mathrm{d}^{3}y\frac{\delta \mu _{0}^{\prime }\left( x^{0},\vec{y}%
\right) }{\delta \pi \left( x\right) }=1,\;  \label{stu16}
\end{equation}
possessing the solution 
\begin{equation}
\mu _{0}^{\prime }\left( y\right) =\pi \left( y\right) ,  \label{stu17}
\end{equation}
that substituted in (\ref{stu14}) reveals the equation $\int \mathrm{d}^{3}y%
\frac{\delta \alpha ^{\prime }\left( x^{0},\vec{y}\right) }{\delta
A^{i}\left( x\right) }=\partial _{i}\varphi \left( x\right) $, clearly
leading to 
\begin{equation}
\alpha ^{\prime }\left( y\right) =A^{i}\partial _{i}\varphi \left( y\right) .
\label{stu18}
\end{equation}
So far, we have generated the first-order deformation of the BRST-invariant
Hamiltonian and BRST charge related to the free model under consideration: 
\begin{equation}
H_{1}=\int \mathrm{d}^{3}x\left( \pi A^{0}+A^{i}\partial _{i}\varphi \right)
,  \label{stu19}
\end{equation}
\begin{equation}
\Omega _{1}=\int \mathrm{d}^{3}x\,\pi \eta ^{2}.  \label{stu20}
\end{equation}
Further, we remark that the first term in the second-order deformation
equation of the BRST-invariant Hamiltonian, (\ref{s10}), is equal to zero
due to $\Omega _{2}=0$; the second piece is found to be 
\begin{equation}
\left[ H_{1},\Omega _{1}\right] =-\int \mathrm{d}^{3}x\left( \partial
_{i}A^{i}\right) \eta ^{2}=\int \mathrm{d}^{3}x\,\rho ^{\prime },
\label{stu21}
\end{equation}
hence (\ref{s10}) is equivalent to $sh_{2}+\rho ^{\prime }=\partial
_{i}k^{\prime i}$, and allows us to write 
\begin{equation}
h_{2}=-\frac{1}{2}A^{i}A_{i},  \label{stu22}
\end{equation}
so that 
\begin{equation}
sh_{2}+\rho ^{\prime }=\partial _{i}\left( -A^{i}\eta ^{2}\right) .
\label{stu23}
\end{equation}
Then, it is easy to check that $\left[ H_{2},\Omega _{1}\right] =0$, which
produces $H_{3}=0$, and consequently $H_{4}=H_{5}=\cdots =0$.

According to the results obtained until now, we can state that the deformed
BRST charge and BRST-invariant Hamiltonian corresponding to the model that
describes an Abelian gauge field coupled with a real scalar field,
consistent to all orders in the deformation parameter, take the form 
\begin{equation}
\Omega =\int \mathrm{d}^{3}x\left( \pi _{0}\eta ^{1}-\left( \partial ^{i}\pi
_{i}-g\pi \right) \eta ^{2}\right) ,  \label{stu24}
\end{equation}
respectively, 
\begin{eqnarray}
&&H_{B}=\int \mathrm{d}^{3}x\left( \frac{1}{2}\pi _{i}\pi _{i}+\frac{1}{4}%
F_{ij}F^{ij}-A^{0}\partial ^{i}\pi _{i}+\frac{1}{2}\pi ^{2}-\frac{1}{2}%
\left( \partial _{i}\varphi \right) \left( \partial ^{i}\varphi \right)
+\right.  \nonumber \\
&&\left. g\pi A^{0}+gA^{i}\partial _{i}\varphi -\frac{1}{2}%
g^{2}A^{i}A_{i}+\eta ^{1}\mathcal{P}_{2}\right) .  \label{stu25}
\end{eqnarray}
On account of these expressions, we deduce that the deformation modifies
only the secondary constraint like 
\begin{equation}
\gamma _{2}^{\prime }\equiv -\partial ^{i}\pi _{i}+g\pi \approx 0,
\label{stu26}
\end{equation}
while the primary one (see the former relation in (\ref{s17})) is unchanged.
In addition, our procedure preserves the Abelianity of the new constraints.
The associated deformed first-class Hamiltonian is 
\begin{eqnarray}
&&H^{\prime }=\int \mathrm{d}^{3}x\left( \frac{1}{2}\pi _{i}\pi _{i}+\frac{1%
}{4}F_{ij}F^{ij}-A^{0}\left( \partial ^{i}\pi _{i}-g\pi \right) +\right. 
\nonumber \\
&&\left. \frac{1}{2}\pi ^{2}-\frac{1}{2}\left( \partial _{i}\varphi
-gA_{i}\right) \left( \partial ^{i}\varphi -gA^{i}\right) \right) .
\label{stu27}
\end{eqnarray}
By passing to the Lagrangian version of the resulting coupled theory, we
find the action 
\begin{equation}
S^{\prime L}\left[ \varphi ,\bar{\varphi},A^{\mu }\right] =\int \mathrm{d}%
^{4}x\left( -\frac{1}{4}F_{\mu \nu }F^{\mu \nu }+\frac{1}{2}\left( \partial
_{\mu }\varphi -gA_{\mu }\right) \left( \partial ^{\mu }\varphi -gA^{\mu
}\right) \right) ,  \label{stu28}
\end{equation}
invariant under the gauge transformations 
\begin{equation}
\delta _{\epsilon }A^{\mu }=\partial ^{\mu }\epsilon ,\;\delta _{\epsilon
}\varphi =g\epsilon .  \label{stu29}
\end{equation}
Thus, the gauge symmetry of the real scalar field in the framework of the
deformed system can again be deduced by performing the gauging of the
corresponding global shift symmetry (\ref{stu2}), present at the level of
the starting free model. Analyzing the coupling between the real scalar
field and the Abelian gauge field emphasized by our deformation procedure,
we conclude that it is precisely a Stuckelberg-like coupling between a zero-
and a one-form.

\section{Couplings between an Abelian gauge field and a Dirac field}

Here, we derive the consistent Hamiltonian interactions between an Abelian
gauge field and a Dirac field, $\left( \psi ^{\alpha },\bar{\psi}_{\alpha
}\right) $. The starting point is a free Lagrangian action that is equal to
the sum between the actions of an Abelian gauge field and a Dirac field 
\begin{equation}
\tilde{S}_{0}^{L}\left[ \psi ^{\alpha },\bar{\psi}_{\alpha },A^{\mu }\right]
=\int \mathrm{d}^{4}x\left( -\frac{1}{4}F_{\mu \nu }F^{\mu \nu }+\bar{\psi}%
_{\alpha }\left( i\left( \gamma ^{\mu }\right) _{\;\;\beta }^{\alpha
}\partial _{\mu }-m\delta _{\;\;\beta }^{\alpha }\right) \psi ^{\beta
}\right) ,  \label{q11}
\end{equation}
where the spinor fields are fermionic, and $\gamma ^{\mu }$ is the standard
notation for Dirac's gamma matrices. The bar operation now signifies Dirac
conjugation. The action (\ref{q11}) is known to be invariant under the
(bosonic) rigid one-parameter symmetry 
\begin{equation}
\Delta \psi ^{\alpha }=i\psi ^{\alpha }\xi ,\;\Delta \bar{\psi}_{\alpha }=-i%
\bar{\psi}_{\alpha }\xi ,  \label{q12}
\end{equation}
involving only the spinors, that gives, according to Noether's theorem, the
conservation law 
\begin{equation}
\partial _{\mu }j^{\mu }=i\frac{\delta ^{R}\mathcal{L}^{\left( D\right) }}{%
\delta \psi ^{\alpha }}\psi ^{\alpha }-i\frac{\delta ^{R}\mathcal{L}^{\left(
D\right) }}{\delta \bar{\psi}_{\alpha }}\bar{\psi}_{\alpha },  \label{q13}
\end{equation}
which emphasizes the conserved current 
\begin{equation}
j^{\mu }=\bar{\psi}_{\alpha }\left( \gamma ^{\mu }\right) _{\;\;\beta
}^{\alpha }\psi ^{\beta },  \label{q14}
\end{equation}
where 
\begin{equation}
\frac{\delta ^{R}\mathcal{L}^{\left( D\right) }}{\delta \psi ^{\alpha }}%
=-\left( i\left( \gamma ^{\mu }\right) _{\;\;\alpha }^{\beta }\partial _{\mu
}+m\delta _{\;\;\alpha }^{\beta }\right) \bar{\psi}_{\beta },  \label{q15}
\end{equation}
\begin{equation}
\frac{\delta ^{R}\mathcal{L}^{\left( D\right) }}{\delta \bar{\psi}_{\alpha }}%
=-\left( i\left( \gamma ^{\mu }\right) _{\;\;\beta }^{\alpha }\partial _{\mu
}-m\delta _{\;\;\beta }^{\alpha }\right) \psi ^{\beta },  \label{q16}
\end{equation}
and $\mathcal{L}^{\left( D\right) }$ obviously denotes the Dirac Lagrangian.
The upper index $R$ ($L$) signifies the right (left) derivative.

From the canonical analysis of this model we extract the constraints and the
canonical Hamiltonian 
\begin{equation}
G_{1}\equiv \pi _{0}\approx 0,\;G_{2}\equiv -\partial ^{i}\pi _{i}\approx 0,
\label{q17}
\end{equation}
\begin{equation}
\chi _{\alpha }\equiv \Pi _{\alpha }-\frac{i}{2}\left( \gamma ^{0}\right)
_{\;\;\alpha }^{\beta }\bar{\psi}_{\beta }\approx 0,\;\bar{\chi}^{\alpha
}\equiv \bar{\Pi}^{\alpha }-\frac{i}{2}\left( \gamma ^{0}\right) _{\;\;\beta
}^{\alpha }\psi ^{\beta }\approx 0,  \label{q18}
\end{equation}
\begin{equation}
\tilde{H}_{0}=\int \mathrm{d}^{3}x\left( \frac{1}{2}\pi _{i}\pi _{i}+\frac{1%
}{4}F_{ij}F^{ij}-A^{0}\partial ^{i}\pi _{i}-\bar{\psi}_{\alpha }\left(
i\left( \gamma ^{i}\right) _{\;\;\beta }^{\alpha }\partial _{i}-m\delta
_{\;\;\beta }^{\alpha }\right) \psi ^{\beta }\right) .  \label{q19}
\end{equation}
In (\ref{q18}--\ref{q19}), $\bar{\Pi}^{\alpha }$ and $\Pi _{\alpha }$ denote
the canonical momenta respectively conjugated to the fields $\bar{\psi}%
_{\alpha }$ and $\psi ^{\alpha }$. The constraints (\ref{q17}) are
first-class and Abelian, while those of (\ref{q18}) are second-class.
Eliminating the second-class constraints by means of the Dirac bracket $%
\left[ ,\right] $ constructed with respect to themselves, we find that the
spinors $\psi ^{\alpha }$ and $\bar{\psi}_{\alpha }$ become conjugated 
\begin{equation}
\left[ \psi ^{\alpha },\bar{\psi}_{\beta }\right] =i\left( \gamma
^{0}\right) _{\;\;\beta }^{\alpha },  \label{q20}
\end{equation}
the resulting theory evolving on a reduced phase-space described by the
fields/momenta $\left( A^{\mu },\pi _{\mu }\right) $, $\left( \psi ^{\alpha
},\bar{\psi}_{\alpha }\right) $ and displaying only the Abelian first-class
constraints (\ref{q17}), together with the first-class Hamiltonian (\ref{q19}%
). Related to the Hamiltonian BRST symmetry associated with this free
theory, we mention that our discussion from section 3.1 remains valid in the
Abelian gauge field sector, with the exception of the bracket, which should
be interpreted as Dirac instead of Poisson. Thus, all formulas (\ref{s21}--%
\ref{s26}) connected with this sector will be used in the sequel, while the
ones describing the complex scalar component should be removed and replaced
by 
\begin{equation}
\mathrm{antigh}\left( \psi ^{\alpha }\right) =\mathrm{antigh}\left( \bar{\psi%
}_{\alpha }\right) =0,\;\mathrm{pgh}\left( \psi ^{\alpha }\right) =\mathrm{%
pgh}\left( \bar{\psi}_{\alpha }\right) =0,  \label{q20a}
\end{equation}
\begin{equation}
\delta \psi ^{\alpha }=0,\;\delta \bar{\psi}_{\alpha }=0,\;\gamma \psi
^{\alpha }=0,\;\gamma \bar{\psi}_{\alpha }=0.  \label{q20b}
\end{equation}
With these observations at hand, we next proceed to analyzing the
Hamiltonian deformation procedure.

The analysis of (\ref{s4}--\ref{s5}), \textit{etc.}, correlated with the
deformation of the BRST charge (\ref{s21}) goes along exactly the same line
as employed for the complex or real scalar field, and allows us to write
down the deformed solution in the form 
\begin{equation}
\Omega =\int \mathrm{d}^{3}x\left( \pi _{0}\eta ^{1}-\left( \partial ^{i}\pi
_{i}-g\tilde{\mu}_{0}\left( \psi ^{\alpha },\bar{\psi}_{\alpha }\right)
\right) \eta ^{2}\right) ,  \label{q31}
\end{equation}
where the (so far) unknown bosonic function $\tilde{\mu}_{0}\left( \psi
^{\alpha },\bar{\psi}_{\alpha }\right) $ depends only on the spinor fields,
is bosonic, and satisfies the properties $\mathrm{antigh}\left( \tilde{\mu}%
_{0}\right) =0$, $\mathrm{pgh}\left( \tilde{\mu}_{0}\right) =0$ and $\gamma 
\tilde{\mu}_{0}=0$. Thus, the only non-vanishing piece in the deformed BRST
charge is the one corresponding to the first-order in the deformation
parameter 
\begin{equation}
\Omega _{1}=\int \mathrm{d}^{3}x\,\tilde{\mu}_{0}\left( \psi ^{\alpha },\bar{%
\psi}_{\alpha }\right) \eta ^{2}.  \label{q31a}
\end{equation}
The unknown function will be found during the identification of the deformed
BRST-invariant Hamiltonian, governed by (\ref{s9}--\ref{s10}), \textit{etc.}

As the BRST-invariant Hamiltonian of the free system under study is (\ref
{s32}), with $H_{0}$ substituted with $\tilde{H}_{0}$, from (\ref{q31a}) it
follows that 
\begin{eqnarray}
&&\left[ H_{0B},\Omega _{1}\right] =-\int \mathrm{d}^{3}x\left( i\left(
i\left( \gamma ^{j}\right) _{\;\;\alpha }^{\beta }\partial _{j}\bar{\psi}%
_{\beta }+m\bar{\psi}_{\alpha }\right) \left( \gamma ^{0}\right) _{\;\;\rho
}^{\alpha }\bar{w}^{\rho }\eta ^{2}+\right.  \nonumber \\
&&\left. i\left( i\left( \gamma ^{j}\right) _{\;\;\alpha }^{\beta }\partial
_{j}\psi ^{\alpha }-m\psi ^{\beta }\right) \left( \gamma ^{0}\right)
_{\;\;\beta }^{\rho }w_{\rho }\eta ^{2}+\eta ^{1}\tilde{\mu}_{0}\right)
=\int \mathrm{d}^{3}x\,\tilde{\lambda},  \label{q33}
\end{eqnarray}
hence (\ref{s9}) reduces in the local form to 
\begin{equation}
sh_{1}+\tilde{\lambda}=\partial ^{i}\tilde{n}_{i},  \label{q34}
\end{equation}
for some $\tilde{n}_{i}$. In (\ref{q33}) we performed the notations $\bar{w}%
^{\rho }\left( x\right) =\int \mathrm{d}^{3}y\frac{\delta ^{L}\tilde{\mu}%
_{0}\left( x^{0},\vec{y}\right) }{\delta \bar{\psi}_{\rho }\left( x\right) }$
and $w_{\rho }\left( x\right) =\int \mathrm{d}^{3}y\frac{\delta ^{L}\tilde{%
\mu}_{0}\left( x^{0},\vec{y}\right) }{\delta \psi ^{\rho }\left( x\right) }$%
. In order to remove the term linear in $\eta ^{1}$ from the left hand-side
of (\ref{q34}), we act like in the case of the complex or real scalar field,
namely, we demand that 
\begin{equation}
h_{1}=\tilde{\mu}_{0}A^{0}+\tilde{\alpha},  \label{q35}
\end{equation}
where the bosonic function $\tilde{\alpha}$ is unknown, and can depend only
on $\psi ^{\alpha }$, $\bar{\psi}_{\alpha }$ and $A^{i}$. The dependence on $%
A^{i}$ is required for ensuring the appearance of spatial derivatives via
the Dirac bracket between $H_{1}$ and $\Omega _{0}$, and, meanwhile, $\tilde{%
\alpha}$ should involve no ghosts or antighosts in order to prevent the
existence of terms in $\left[ H_{1},\Omega _{0}\right] $ different from
those in $\left[ H_{0B},\Omega _{1}\right] $, which can be attained via a
dependence also on $\psi ^{\alpha }$ and $\bar{\psi}_{\alpha }$.
Accordingly, we find 
\begin{equation}
\left[ H_{1},\Omega _{0}\right] =\int \mathrm{d}^{3}x\left( \eta ^{1}\tilde{%
\mu}_{0}+\tilde{a}_{i}\partial ^{i}\eta ^{2}\right) ,  \label{q36}
\end{equation}
where $\tilde{a}_{i}\left( x\right) =\int \mathrm{d}^{3}y\frac{\delta \tilde{%
\alpha}\left( x^{0},\vec{y}\right) }{\delta A^{i}\left( x\right) }$. From (%
\ref{q33}) and (\ref{q36}), we see that (\ref{q34}) becomes 
\begin{eqnarray}
&&-i\left( \left( i\left( \gamma ^{j}\right) _{\;\;\alpha }^{\beta }\partial
_{j}\bar{\psi}_{\beta }+m\bar{\psi}_{\alpha }\right) \left( \gamma
^{0}\right) _{\;\;\rho }^{\alpha }\bar{w}^{\rho }\right.  \nonumber \\
&&\left. +\left( i\left( \gamma ^{j}\right) _{\;\;\alpha }^{\beta }\partial
_{j}\psi ^{\alpha }-m\psi ^{\beta }\right) \left( \gamma ^{0}\right)
_{\;\;\beta }^{\rho }w_{\rho }\right) \eta ^{2}+\tilde{a}_{i}\partial
^{i}\eta ^{2}=\partial ^{i}\tilde{n}_{i}.  \label{q37}
\end{eqnarray}
The left hand-side of (\ref{q37}) reduces to a total derivative if 
\begin{eqnarray}
&&-i\left( \left( i\left( \gamma ^{j}\right) _{\;\;\alpha }^{\beta }\partial
_{j}\bar{\psi}_{\beta }+m\bar{\psi}_{\alpha }\right) \left( \gamma
^{0}\right) _{\;\;\rho }^{\alpha }\bar{w}^{\rho }\right.  \nonumber \\
&&\left. +\left( i\left( \gamma ^{j}\right) _{\;\;\alpha }^{\beta }\partial
_{j}\psi ^{\alpha }-m\psi ^{\beta }\right) \left( \gamma ^{0}\right)
_{\;\;\beta }^{\rho }w_{\rho }\right) =\partial ^{i}\tilde{a}_{i}.
\label{q38}
\end{eqnarray}
Adding to both hand-sides of (\ref{q38}) the term $-\left( i\left( \gamma
^{0}\right) _{\;\;\alpha }^{\beta }\partial _{0}\bar{\psi}_{\beta }\right)
i\left( \gamma ^{0}\right) _{\;\;\rho }^{\alpha }\bar{w}^{\rho }$, as well
as the quantity $-i\left( \left( \gamma ^{0}\right) _{\;\;\alpha }^{\beta
}\partial _{0}\psi ^{\alpha }\right) i\left( \gamma ^{0}\right) _{\;\;\beta
}^{\rho }w_{\rho }$, we deduce 
\begin{eqnarray}
&&\frac{\delta ^{R}\mathcal{L}^{\left( D\right) }}{\delta \psi ^{\alpha }}%
i\left( \gamma ^{0}\right) _{\;\;\rho }^{\alpha }\bar{w}^{\rho }+\frac{%
\delta ^{R}\mathcal{L}^{\left( D\right) }}{\delta \bar{\psi}_{\alpha }}%
i\left( \gamma ^{0}\right) _{\;\;\alpha }^{\rho }w_{\rho }=  \nonumber \\
&&\left( \partial _{0}\bar{\psi}_{\alpha }\right) \bar{w}^{\alpha }+\left(
\partial _{0}\psi ^{\alpha }\right) w_{\alpha }+\partial ^{i}\tilde{a}_{i}.
\label{q39}
\end{eqnarray}
Analyzing the structure of the last formula and replacing $\bar{w}^{\rho }$, 
$w_{\rho }$ in terms of $\tilde{\mu}_{0}$, it turns out that its left
hand-side gives the variation of the Dirac Lagrangian under the rigid
one-parameter transformations 
\begin{equation}
\Delta \psi ^{\alpha }\left( x\right) =i\left( \gamma ^{0}\right) _{\;\;\rho
}^{\alpha }\int \mathrm{d}^{3}y\frac{\delta ^{L}\tilde{\mu}_{0}\left( x^{0},%
\vec{y}\right) }{\delta \bar{\psi}_{\rho }\left( x\right) }\xi ,
\label{q40a}
\end{equation}
\begin{equation}
\Delta \bar{\psi}_{\alpha }\left( x\right) =i\left( \gamma ^{0}\right)
_{\;\;\alpha }^{\rho }\int \mathrm{d}^{3}y\frac{\delta ^{L}\tilde{\mu}%
_{0}\left( x^{0},\vec{y}\right) }{\delta \psi ^{\rho }\left( x\right) }\xi .
\label{q40b}
\end{equation}
Identifying (\ref{q40a}) and (\ref{q40b}) with the well-known global
one-parameter invariance (\ref{q12}) of Dirac theory, we are led to the
equations 
\begin{equation}
\left( \gamma ^{0}\right) _{\;\;\rho }^{\alpha }\int \mathrm{d}^{3}y\frac{%
\delta ^{L}\tilde{\mu}_{0}\left( x^{0},\vec{y}\right) }{\delta \bar{\psi}%
_{\rho }\left( x\right) }=\psi ^{\alpha }\left( x\right) ,  \label{q41a}
\end{equation}
\begin{equation}
\left( \gamma ^{0}\right) _{\;\;\alpha }^{\rho }\int \mathrm{d}^{3}y\frac{%
\delta ^{L}\tilde{\mu}_{0}\left( x^{0},\vec{y}\right) }{\delta \psi ^{\rho
}\left( x\right) }=-\bar{\psi}_{\alpha }\left( x\right) ,  \label{q41b}
\end{equation}
that yield the solution 
\begin{equation}
\tilde{\mu}_{0}\left( y\right) =\bar{\psi}_{\alpha }\left( y\right) \left(
\gamma ^{0}\right) _{\;\;\beta }^{\alpha }\psi ^{\beta }\left( y\right) .
\label{q42}
\end{equation}
Substituting (\ref{q42}) in (\ref{q39}) and using (\ref{q13}), we are
provided with the equations $\int \mathrm{d}^{3}y\frac{\delta \tilde{\alpha}%
\left( x^{0},\vec{y}\right) }{\delta A_{i}\left( x\right) }=\bar{\psi}%
_{\alpha }\left( x\right) \left( \gamma ^{i}\right) _{\;\;\beta }^{\alpha
}\psi ^{\beta }\left( x\right) $, which produce 
\begin{equation}
\tilde{\alpha}\left( y\right) =\bar{\psi}_{\alpha }\left( y\right) \left(
\gamma ^{i}\right) _{\;\;\beta }^{\alpha }\psi ^{\beta }\left( y\right)
A_{i}\left( y\right) .  \label{q43}
\end{equation}
Consequently, we have generated the first-order deformed BRST-invariant
Hamiltonian: 
\begin{equation}
H_{1}=\int \mathrm{d}^{3}x\left( \bar{\psi}_{\alpha }\left( \gamma
^{0}\right) _{\;\;\beta }^{\alpha }\psi ^{\beta }A_{0}+\bar{\psi}_{\alpha
}\left( \gamma ^{i}\right) _{\;\;\beta }^{\alpha }\psi ^{\beta }A_{i}\right)
.  \label{q44}
\end{equation}
Further, let us study the higher-order deformations. By direct computation
we get $\left[ H_{1},\Omega _{1}\right] =0$, which combined with $\Omega
_{2}=0$ allows us to take the solution of (\ref{s10}) to be $H_{2}=0$. Then,
it is simply to check that we can choose 
\begin{equation}
H_{3}=H_{4}=\cdots =0.  \label{q45}
\end{equation}

In conclusion, the complete deformed BRST charge and BRST-invariant
Hamiltonian that govern the couplings between an Abelian gauge field and the
Dirac field are given by 
\begin{equation}
\Omega =\int \mathrm{d}^{3}x\left( \pi _{0}\eta ^{1}-\left( \partial ^{i}\pi
_{i}-g\bar{\psi}_{\alpha }\left( \gamma ^{0}\right) _{\;\;\beta }^{\alpha
}\psi ^{\beta }\right) \eta ^{2}\right) ,  \label{q46}
\end{equation}
respectively, 
\begin{eqnarray}
&&H_{B}=\int \mathrm{d}^{3}x\left( \frac{1}{2}\pi _{i}\pi _{i}+\frac{1}{4}%
F_{ij}F^{ij}-A^{0}\partial ^{i}\pi _{i}-\right.  \nonumber \\
&&\left. \bar{\psi}_{\alpha }\left( i\left( \gamma ^{i}\right) _{\;\;\beta
}^{\alpha }\partial _{i}-m\delta _{\;\;\beta }^{\alpha }\right) \psi ^{\beta
}+g\bar{\psi}_{\alpha }\left( \gamma ^{\mu }\right) _{\;\;\beta }^{\alpha
}\psi ^{\beta }A_{\mu }+\eta ^{1}\mathcal{P}_{2}\right) .  \label{q47}
\end{eqnarray}
Like in the case of the scalar field theory, from the above quantities we
read off that the classical Hamiltonian interacting theory is subject to the
deformed Abelian first-class constraints 
\begin{equation}
\tilde{\gamma}_{2}\equiv -\partial ^{i}\pi _{i}+g\bar{\psi}_{\alpha }\left(
\gamma ^{0}\right) _{\;\;\beta }^{\alpha }\psi ^{\beta }\approx 0,
\label{q48}
\end{equation}
and the former constraint in (\ref{q17}), as well as that the first-class
Hamiltonian with respect to these constraints has the expression 
\begin{eqnarray}
&&\tilde{H}=\int \mathrm{d}^{3}x\left( \frac{1}{2}\pi _{i}\pi _{i}+\frac{1}{4%
}F_{ij}F^{ij}-A^{0}\left( \partial ^{i}\pi _{i}-g\bar{\psi}_{\alpha }\left(
\gamma ^{0}\right) _{\;\;\beta }^{\alpha }\psi ^{\beta }\right) -\right. 
\nonumber \\
&&\left. \bar{\psi}_{\alpha }\left( i\left( \gamma ^{i}\right) _{\;\;\beta
}^{\alpha }\partial _{i}-m\delta _{\;\;\beta }^{\alpha }\right) \psi ^{\beta
}+g\bar{\psi}_{\alpha }\left( \gamma ^{\mu }\right) _{\;\;\beta }^{\alpha
}\psi ^{\beta }A_{\mu }\right) ,  \label{q49}
\end{eqnarray}
where the first-class behaviour is considered in terms of the Dirac bracket (%
\ref{q20}). If we take the necessary steps to the Lagrangian framework, we
discover that the resulting interacting theory displays the Lagrangian
action 
\begin{equation}
\tilde{S}^{L}\left[ \psi ^{\alpha },\bar{\psi}_{\alpha },A^{\mu }\right]
=\int \mathrm{d}^{4}x\left( -\frac{1}{4}F_{\mu \nu }F^{\mu \nu }+\bar{\psi}%
_{\alpha }\left( i\left( \gamma ^{\mu }\right) _{\;\;\beta }^{\alpha }{D}%
_{\mu }-m\delta _{\;\;\beta }^{\alpha }\right) \psi ^{\beta }\right) ,
\label{q50}
\end{equation}
invariant under the gauge transformations 
\begin{equation}
\delta _{\epsilon }A^{\mu }=\partial ^{\mu }\epsilon ,\;\delta _{\epsilon
}\psi ^{\alpha }=ig\psi ^{\alpha }\epsilon ,\;\delta _{\epsilon }\bar{\psi}%
_{\alpha }=-ig\bar{\psi}_{\alpha }\epsilon ,  \label{q51}
\end{equation}
where the covariant derivative ${D}_{\mu }$ takes the form (\ref{s53}). We
observe that, exactly like for the complex or real scalar field, the spinors
bear now some gauge invariances, resulting from the original rigid ones in a
direct manner by gauging. An interesting difference between this model and
the scalar theory is that while there we have obtained nontrivial pieces for
the BRST-invariant Hamiltonian at order two in the deformation parameter,
the similar quantity stops here at order one. This feature is essentially
due to the statistics of the present matter fields, which are spinors, hence
fermionic. Thus, we can conclude that as a result of our deformation scheme
we obtained the well-known model describing the coupling between the
electromagnetic and spinor fields, namely, spinor electrodynamics.

\section{Conclusion}

In conclusion, in this paper we have derived the consistent Hamiltonian
interactions between an Abelian gauge field and the complex scalar field,
the massless real scalar field, respectively, Dirac field. Our approach is
based on the deformation of the BRST charge and BRST-invariant Hamiltonian
associated with the uncoupled theories involving these fields. The
derivation of the solutions to the main equations that govern our BRST
deformation procedure essentially relies on the presence of some conserved
currents corresponding to the rigid symmetries of the matter fields from the
``free'' models. The first-order deformations of both BRST charge and
BRST-invariant Hamiltonian can be written in the form $\Omega _{1}=\pm \int 
\mathrm{d}^{3}xq\eta ^{2}$, respectively, $H_{1}=\pm \int \mathrm{d}%
^{3}x\left( qA^{0}+j_{i}A^{i}\right) $, in the case of all analysed models,
where $q$ is the Hamiltonian charge density of the associated conserved
currents. For the scalar case we have that $\left[ H_{1},\Omega _{1}\right] $
is non-vanishing due to the fact that $\left[ j_{i},q\right] $ is not zero,
which requires non-trivial second-order deformations of the BRST-invariant
Hamiltonian. In the case of the Dirac theory we have $\left[ H_{1},\Omega
_{1}\right] =0$, so the second-order deformations of $H_{0B}$ can be taken
to vanish. It is interesting to note that, apart from others situations~\cite
{39,40,41,42}, where the deformation of the BRST charge can be computed in a
self-consistent manner, here we need to alternate it with the deformation of
the BRST-invariant Hamiltonian in order to reach some complete solutions. As
a result of our method we discover scalar electrodynamics, a
Stuckelberg-like coupling, respectively, spinor electrodynamics. All the
couplings are local, and the matter fields bear some gauge invariances that
can be produced via the gauging of the original global symmetries. As
expected, the $U(1)$ gauge invariance of Maxwell's field is kept unchanged
for all models during the deformation process.

\section*{Acknowledgment}

This work has been supported by a Romanian National Council for Academic
Scientific Research (CNCSIS) grant.

\end{document}